# Highly asymmetric interaction forces induced by acoustic waves in coupled plate structures

Xiying Fan,[1] Chunyin Qiu,[1] Shenwei Zhang,[1] Manzhu Ke,[1] and Zhengyou Liu[1,2a]

[1]Key Laboratory of Artificial Micro- and Nano-structures of Ministry of Education and School of Physics and Technology, Wuhan University, Wuhan 430072, China

[2]Institute for Advanced Studies, Wuhan University, Wuhan 430072, China

**Abstract:** Mutual forces can be induced between coupled structures when illuminated by external acoustic waves. In this Letter, we propose a concept of asymmetric interaction between two coupled plate-like structures, which is generated by oppositely incident plane waves. Besides the striking contrast in magnitude, the mutual force induced by one of the incidences can be tuned extremely strong due to the resonant excitation of the flexural plate modes. The highly asymmetric interaction with enhanced strength in single side should be potentially useful, such as in designing ultrasound instruments and sensors.

[a]Author to whom correspondence should be addressed. Email: zyliu@whu.edu.cn

Recently, the asymmetric or unidirectional acoustic transmission has aroused wide interest [1-8] because of its prospective applications, such as in designing acoustic sensors and rectifying devices. The asymmetric transmission can be realized by breaking either the time-reversal symmetry or spatial-inversion symmetry along the opposite incident directions. For the former case, the symmetry breaking can be obtained by introducing nonlinear media [1,2] or macroscopic flow field [3,4]. For the latter case, the design is relatively simple: principally, the asymmetric sound transmission is produced by the mode conversion of multiple scattering channels. A representative example can be referred to the spatially asymmetric gratings, as long as the nonzero-order of diffraction is involved. In this case, highly asymmetric sound transmission has been realized frequently [5-8].

In this Letter, we propose a concept of asymmetric interaction between two coupled plate structures, which is induced by acoustic plane wave incident from opposite directions. It is well-known that acoustic wave can exert acoustic radiation force (ARF) on the object illuminated [9], where the characteristic of ARF depends on the properties of the object and sound source simultaneously. In recent decades, various manipulations of sound fields have been realized by tailoring artificial structures, e.g. negative refraction [10-12], directional radiation [13-16], and wavefront reshaping [17-19]. The exotic sound fields, in turn, could be used to produce abnormal ARFs. Recently, the capability of particle trapping and sorting has been well demonstrated by using the strongly localized gradient fields generated by patterned plates [20-23]. For coupled artificial structures, a rich diversity of acoustically induced mutual forces (AIMFs) between the monomers have also been observed [24,25], where the AIMF magnitudes can be tailored much larger than the ARFs acting on the whole systems. The property of AIMF is closely related to the vibration morphology of resonant field. In air ambience, flexibly tunable repulsive forces have been firstly realized between a metamaterial slab and a rigid wall by introducing cavity resonances [24]. Furthermore, for a pair of water-immersed structured plates [25], strong attractions and repulsions have been achieved simultaneously, due to the resonant excitation of the coupled flexural modes in plates.

In fact, the asymmetric AIMF can be widely observed in asymmetric coupled systems. As an extreme example, the interaction between the metamaterial slab and rigid wall is definitely highly asymmetric [24]: for the incidence from the metamaterial slab, the interaction can be tuned considerably strong (with respect to the ARF acting on the slab or rigid wall alone), whereas for the incidence from the impenetrable wall, there is no interaction induced by the acoustic wave. Usually, the rigid approximation takes effect for gas ambience. However, for liquid background (e.g., water, one of the most frequently encountered sound environments), the elasticity of solid must be considered carefully. In this paper, we have designed a water-immersed sample made of a couple of asymmetric steel plate structures, which exhibits strong interaction for one incidence, together with striking magnitude contrast to that of opposite incidence. Different from the asymmetric transmission, the information contrast can be manifested through mechanic force directly (without measuring sound field). Moreover, for the coupled plate structures, the asymmetry can occur even in the case of the symmetric transmission, since the AIMF depends on the near-field distribution only. Potential applications can be conceived based this phenomenon, such as to design hydroacoustic sensors.

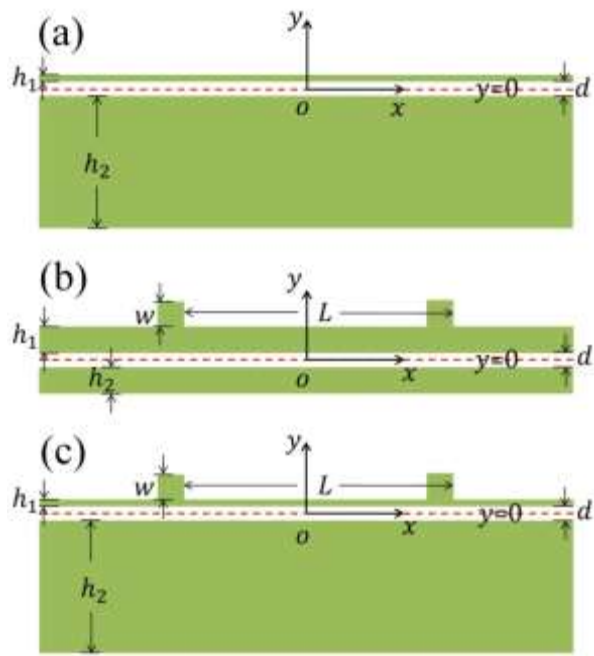

FIG. 1. (color online). Schematic illustrations for the asymmetric structures under consideration, each made of a couple of parallel steel plates immersed in water. (a) Sample-1: two planar plates of different thicknesses. (b) Sample-2: a planar plate plus a patterned plate with identical thickness. (c) Sample-3: the same as sample-1, but with square bumps patterned on the thin plate. The horizontal dashed line ($y=0$) indicates the integral surface for evaluating the AIMF between plates.

As illustrated in Fig. 1, in this paper three typical double-plate structures are considered, which are immersed in water and placed in parallel at a distance $d$. For the periodical structures, we focus on the frequency range below the first-order diffraction, i.e. $L/\lambda_0 < 1$, where $L$ is the structure period and $\lambda_0$ is the wavelength in water. In this case, although the inversion symmetry of sample is broken, the power transmission is symmetric due to the reciprocity. The material parameters used are listed below: mass density $\rho_s = 7.67 g/cm^3$, longitudinal velocity $c_l = 6.01 km/s$ and transverse velocity $c_t = 3.23 km/s$ for steel; mass density $\rho_0 = 1.0 g/cm^3$ and sound velocity $c_0 = 1.49 km/s$ for water.

As stated in Ref. [25], for the incidence along +y direction, the dimensionless AIMF between plates can be simply written as a surface integral at $y=0$,

$$Y_m = \frac{1}{2L}\int_{-L/2}^{L/2} \left(|p/p_0|^2 + |u_y/u_0|^2 - |u_x/u_0|^2\right)dx, \tag{1}$$

where the pressure $p$ and velocities $u_x$ and $u_y$, excited by the plane wave with pressure and velocity amplitudes $p_0$ and $u_0 = p_0/(\rho_0 c_0)$, can be evaluated numerically by the finite-element solver (COMSOL MULTIPHYSICS). The AIMF is scaled by the unit $E_0 S$, where $E_0$ is the energy density of the incident plane wave, and $S$ is the total area of the planar plate. The positive and negative signs of $Y_m$ correspond to repulsive and attractive interactions, respectively. Similar formula can be written for the incidence along –y direction. Since the dimensionless value of AIMF could span over several orders, for clarity hereafter we show $\log_{10}$-plot of the AIMF magnitude $F^s = |Y_m|$, where the sign $s = \pm$ denotes the incidence along $\pm y$ direction. The definition of AIMF is particularly meaningful if $F^s \gg 1$ (occurring near resonance), since it can be approximated to the (directly measurable) ARF acting on each monomer. It is worth pointing out that, since the AIMF depends on the near-field distribution inside the gap between plates, it can be tuned highly asymmetric without breaking any fundamental law. This is in contrast to the ARF acting on the whole coupled system (with a maximum value of 2 in total reflection case), which can be evaluated from the far-field distribution and thus invariant to the incidence due to the reciprocity.

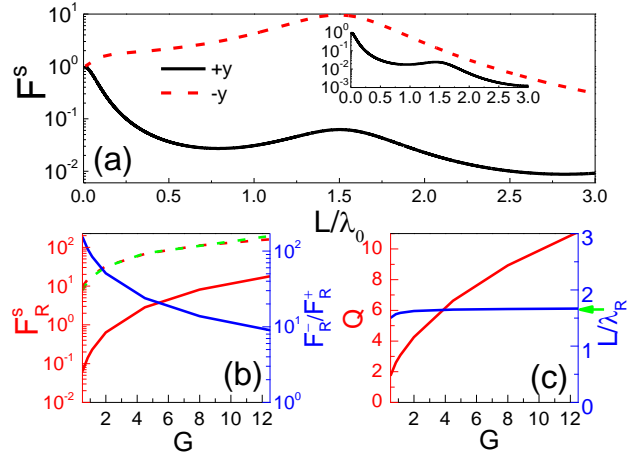

FIG. 2 (color online). (a) The AIMF spectra for the sample-1, produced by a plane wave incident from $+y$ or $-y$ direction, where the inset shows the incidence independent power transmission spectrum. (b) The AIMFs $F_R^+$ (red solid) and $F_R^-$ (red dashed), and their ratios $F_R^-/F_R^+$ (blue solid) for the systems with different geometric ratios $G = h_1/d$, extracted from the resonant peaks of the AIMF spectra. For comparison, the theoretical prediction of $F_R^-$ (green dashed) is also presented. (c) The $G$-dependent $Q$-factor (red solid) and resonant frequency (blue solid) of the AIMF peak, where the green arrow indicates the predicted resonant frequency. Both quantities are verified to be insensitive to the incidence.

As shown in Fig. 1(a), the sample-1 consists of a couple of planar plates with remarkably different thicknesses, $h_1 = L/40$ and $h_2 = L/2$, placed at a distance $d = L/20$. For this sample, it is intuitively that there is a strong contrast between the AIMFs $F^+$ and $F^-$, induced by the plane waves from +y and −y directions. In Fig. 2(a) we present the $F^s$ spectra, together with the normalized transmission for reference (see inset). It is observed that as the dimensionless frequency $L/\lambda_0$ goes to zero, both $F^+$ and $F^-$ approach to the value of 1, as a consequence of full penetrability, i.e. $|p|=|p_0|$ and $|u_y|=|u_0|$ ($|u_x|\equiv 0$). As $L/\lambda_0$ increases, as expected by intuition, $F^-$ becomes much larger than $F^+$, since the acoustic field can easily penetrate the thin plate to form stronger wave concentration inside the gap between plates. In practical applications, a big value of $F^-$ is highly desirable. Consistent with the soft peak in transmission, both AIMF spectra show maximized values at $L/\lambda_0 = 1.51$, $F^- \approx 9.59$ and $F^+ \approx 0.062$, associated with a high contrast ratio $F^-/F^+ \approx 155$. The further study states that at this frequency the normalized pressure field $|p/p_0|$ is almost uniform inside the gap and dominant over the normalized velocity field $|u_y/u_0|$, which resembles the cavity resonance reported in Ref. [24] since the lower plate is nearly at rest. According to a similar derivation, the resonant frequency can be analytically estimated from the formula $L/\lambda_R = L/\left[2\pi\sqrt{h_1 d \rho_s / \rho_0}\right]$, and the AIMF induced from the penetrable thin plate can be approximated to $F_R^- = 2(\rho_s/\rho_0)(h_1/d)$.

The analytical formula states that $F_R^-$ is proportional to the density ratio $D = \rho_s/\rho_0$ and the geometric ratio $G = h_1/d$. For the water background, there is little space to increase $D$ and the further enhancement of $F_R^-$ should be resorted to the tunable factor $G$. This can be seen from the $G$-dependence of $F_R^-$ in Fig. 2(b), where the red and green dashed lines indicate the (consistent) results from full-wave simulation and theoretical prediction, respectively. Note that here the product $h_1 d$ is fixed to keep the resonant frequency $L/\lambda_R$ nearly invariant. In Fig. 2(b), we also present the corresponding values of $F_R^+$ (red solid line) and $F_R^-/F_R^+$ (blue solid line). It is observed that, despite of the linear growth of $F_R^-$ as $G$, the AIMF ratio $F_R^-/F_R^+$ decreases quickly due to the reduction of the thickness ratio $h_2/h_1$. For example, for the case of $G=4.5$, $F_R^-$ and $F_R^+$ increase to 67.4 and 2.8, respectively, and consequently $F_R^-/F_R^+$ reduces to 23.9. In practical applications, the quality factor of the AIMF peak $Q$, defined as the inverse of the relative frequency width of the half peak maximum, is also an important quantity to characterize the performance of system. It is observed in Fig. 2(c) that the $Q$-factor, verified to be insensitive to the incidence, increases from the order of $10^0$ to $10^1$. This tendency is consistent with the growing $F^-$: as the plate-plate gap is narrowed down (to increase $G$), the pressure field in between turns stronger and the resonance gets sharper. Furthermore, in Fig. 2(c) we also present the numerical resonant frequency (blue solid line), which is predicted well by the above formula (see green arrow) if only the upper plate is not too thin.

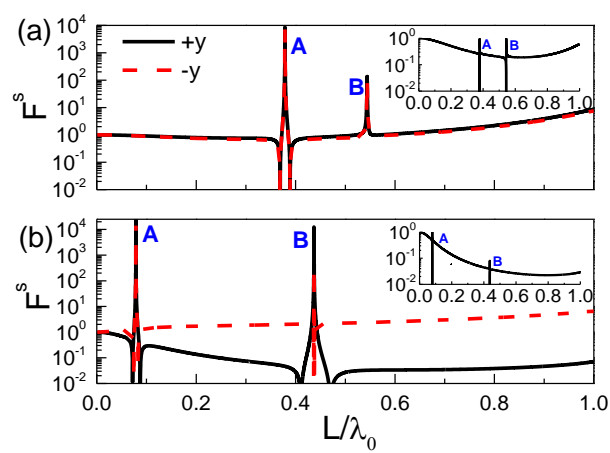

FIG. 3 (color online). The AIMF spectra for the sample-2 (a) and sample-3 (b), where the insets show the corresponding transmission spectra. Note that each striking dip in the AIMF spectra implies a crossover between repulsion and attraction.

As stated above, in the planar plate system it is not easy to obtain strong interaction for one incidence, together with striking contrast to the other one simultaneously. Besides, the interaction is always repulsive since $u_x \equiv 0$ at normal incidence [see Eq. (1)]. Recently, it was manifested [25] that the AIMF (either repulsive or attractive) between a pair of identical elastic plates can be greatly enhanced by patterning periodic bumps, because of the resonant excitation of the coupled flexural plate modes. In that case, the system is symmetric with respect to the middle plane $y=0$. To break the symmetry, as shown in Fig. 1(b), in sample-2 only the upper plate is patterned with periodic square bumps. Specifically, here the plate thicknesses $h_1 = h_2 = L/10$, the separation $d = L/20$, and the bump size $w = L/10$. The AIMF spectra for both incidences are presented in Fig. 3(a), which display two striking peaks, depicted by *A* and *B*, consistent with the sharp Fano-resonances indicated in the transmission spectrum (see inset). Similar to Ref. [25], the further study states that these enhanced attraction (peak *A*) or repulsion (peak *B*) can be attributed to the resonant excitation of the anti-phase or in-phase coupling of the flexural plate modes. In spite of the huge magnitudes of the peaked AIMFs, however, there is no significant difference observed between +*y* and −*y* incidences.

Now we turn to the sample-3. For comparison, both plate thicknesses are identical to the sample-1, and the bump size is the same as in sample-2. The sample is expected to have both advantages of the sample-1 and sample-2: the existence of periodical structure strongly enhances the AIMF magnitude, and the large thickness ratio contributes high contrast for opposite incidences. The AIMF spectra are presented in Fig. 3(b), exhibiting two striking peaks again in the considered subwavelength region. For peak *A*, the AIMF magnitudes generated by both incidences are very close, $F_R^+ \approx 23000$ and $F_R^- \approx 21500$, because of the high penetrability of the plates at low frequency. For the peak *B* at higher frequency, however, the contrast is remarkable: the normalized value for the +*y* incidence, $F_R^+ \approx 12300$, is about 33 times to that of −*y* incidence, $F_R^- \approx 370$. The further observation manifests that the AIMFs behave as attractive around both resonances, in contrast to those away from the resonances, which are always repulsive (consistent with the sample-1).

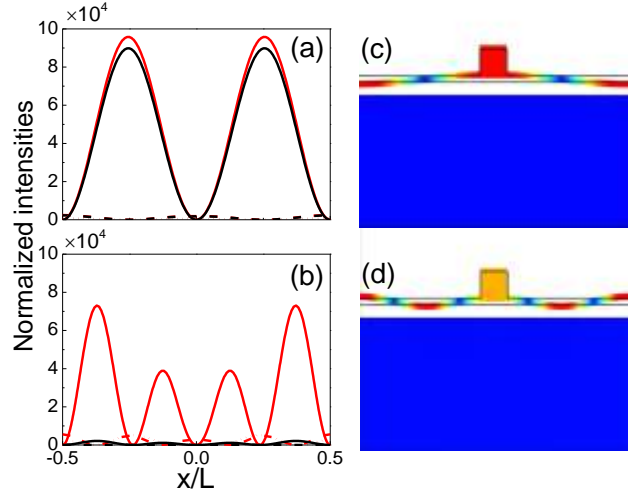

FIG. 4 (color online). The dimensionless intensity quantities measured at $y=0$, i.e., $|p/p_0|^2$ (dotted), $|u_y/u_0|^2$ (dashed), and $|u_x/u_0|^2$ (solid) for the incidences along $+y$ (red) or $-y$ (black) direction, where (a) and (b) for the resonances **A** and **B**, respectively. (c) and (d), the corresponding (instant) vibration morphologies for the resonances excited along $+y$ direction.

To obtain a direct understanding of the strong attraction in the sample-3 at both resonances, in Figs. 4(a) and 4(b) we present the dimensionless intensities for the pressure and velocity fields distributed at $y=0$. It is observed that for both cases, the intensity quantity $|u_x/u_0|^2$ dominates over the other two, $|p/p_0|^2$ and $|u_y/u_0|^2$, which leads to attractive interaction according to Eq. (1). The physics essential can be revealed in Figs. 4(c) and 4(d), which display the corresponding vibration morphologies for $+y$ incidence. Obviously, the vibration is mostly localized in the upper thin plate, which is typically flexural motion. As reported previously [26], the nonleaky flexural mode can be resonantly excited when additional momentum is supplied by the structure periodicity, i.e., $k_p = 2n\pi/L$, where $k_p$ is the wavenumber of flexural mode. Here the resonances **A** and **B** correspond to the $n=1$ and $n=2$, consistent with the prediction from the dispersion relation of the flexural mode. In fact, by considering the less bended thick plate as a mirror, the vibration morphologies for both resonances resemble the anti-phase-coupled flexural modes as depicted in Ref. [25], in which the intensity component $|u_x/u_0|^2$ plays a dominant role in the attractive interaction. [This is different from the sample-2, where both plates bend simultaneously and the in-phase coupling may also contribute repulsive interaction.] Similar vibration morphologies can also be observed in the case of $-y$ incidence.

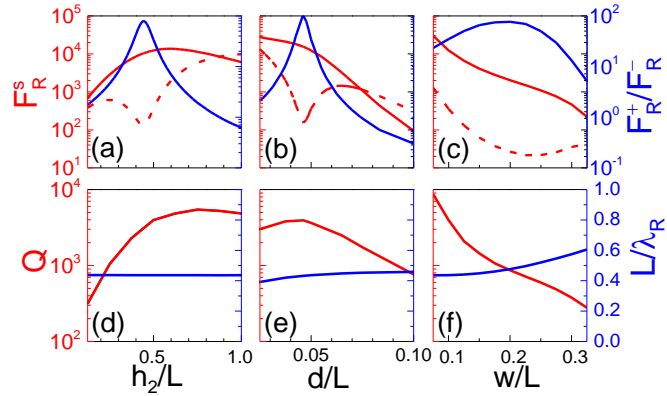

FIG. 5 (color online). (a)-(c) The parameter dependences of the AIMF magnitudes of the peak ***B*** for $+y$ (red solid) and $-y$ (red dashed) incidences, together with their ratios (blue solid). (d)-(f) The quality factors (red solid) and resonant frequencies (blue solid) of the AIMF peak ***B*** varying with the same parameters.

To further study the properties of AIMFs at resonance ***B***, in Figs. 5(a)-5(c), we present the AIMF magnitudes varying with the thickness of the lower plate $h_2$, the plate-plate distance $d$, and the width of the square bump $w$. It is observed that the peaked AIMFs are sensitive to all of these geometric parameters, associated with non-monotonous AIMF ratio $F_R^+/F_R^-$, which allow us to further improve the performance of the system. For example, when $d/L = 0.046$ the AIMF ratio can be optimized to $F_R^+/F_R^- \approx 100$ while keeping high strength of $F_R^+ \approx 15400$, see Fig. 5(b). [For the resonance ***A***, without data presented here, $F_R^+/F_R^-$ always approaches to 1 despite of huge magnitudes for $F_R^+$ and $F_R^-$.] Also, from Figs. 5(a) and 5(b) we can find that $F_R^+$ can be tuned either larger or smaller than $F_R^-$. Therefore, the intuition used for sample-1 (i.e., the stronger AIMF is expected for the incidence from thin plate) fails in this complex structure. In Figs. 5(d)-(f), we present the corresponding parameter dependence of the *Q*-factor for the AIMF peak, which exhibits remarkable sensitivity again. It is of interest that the *Q*-factor is much higher than that of sample-1, consistent with the stronger AIMF amplitude due to the sharp Fano-resonance. In Figs. 5(d)-(f), we also present the resonant frequency for these tunable geometric parameters. It is observed that the resonant frequency is less sensitive to $h_2$ and $d$, since the resonance is determined mostly by the bending motion in the thin plate. In contrast, the resonant frequency is more sensitive to $w$ (growing with $w$ due to the effective shortening of the soft thin plate).

In summary, we have studied the acoustically-induced interaction in some typical asymmetric parallel-plate structures. It is found that the periodical patterning contributes to huge strength of interaction, whereas a large ratio of plate thickness results in remarkable magnitude contrast between opposite incidences. The combination gives a system of high performance: for one incidence, the interaction strength can be tuned four orders higher than that acting on the whole sample, accompanying with a contrast ratio of two orders to that induced by opposite incidence. This exotic effect is caused by the asymmetric excitation of the flexural mode inside the thin plate. Prospective applications of such sensitive asymmetric interaction can be anticipated, e.g. in designing sensors to identify the propagating direction of weak sound. A similar concept can also be introduced to optic systems to miniaturize devices.

**Acknowledgements**

This work is supported by the National Basic Research Program of China (Grant No. 2015CB755500); National Natural Science Foundation of China (Grant Nos. 11174225, 11004155, 11374233, and J1210061).